\documentclass[a4paper]{jpconf}

\usepackage[pdftex]{graphicx}
\usepackage{epstopdf}
\usepackage{psfrag}
\usepackage{amsmath}        
\usepackage{amsfonts}              
\usepackage{amsthm}                
\usepackage{amscd}

\newcommand{\R}{\mathbb{R}}

\newcommand{\norm}[1]{\left\lVert #1 \right\rVert}

\newcommand{\ket}[1]{\lvert\, #1\rangle}

\newcommand{\expec}[1]{\langle #1 \rangle}
\newcommand{\ig}[2]{\includegraphics}

\newcommand{\pb}[1]{\underset{\Leftarrow}{#1}}

\newcommand{\lp}{\ell_\text{p}}

\newcommand{\hilb}{\mathcal{H}}
\newcommand{\nj}{\,\rvert_j}
\newcommand{\njp}{\,\rvert_{j'}}
\newcommand{\njd}{\dashv_{j}}
\newcommand{\dm}{\text{d}\mu}

\newcommand{\hal}{\hilb_{\text{AL}}}
\newcommand{\bg}[1]{#1^{(0)}}
\newcommand{\pin}{\pi'}
\newcommand{\vac}{\text{vac}}
\newcommand{\wt}[1]{\widetilde{#1}}

\DeclareMathOperator{\conn}{\mathcal{A}}
\DeclareMathOperator{\tra}{tr}

\DeclareMathOperator{\aad}{ad}

\DeclareMathOperator{\one}{\boldsymbol{1}}
\DeclareMathOperator{\um}{\mathbb{I}}
\DeclareMathOperator{\id}{id}

\DeclareMathOperator{\hfalg}{\mathfrak{HF}}

\DeclareMathOperator{\pone}{\triangleright}
\DeclareMathOperator{\pont}{\triangleright}

\begin{document}
\title{New insights in quantum geometry}

\author{Hanno Sahlmann}

\address{Asia Pacific Center for Theoretical Physics, Pohang (South Korea)\\
Physics Department, Pohang University of Science and Technology, Pohang (South Korea)}

\ead{sahlmann@apctp.org}

\begin{abstract}
Quantum geometry, i.e., the quantum theory of intrinsic and extrinsic spatial geometry, is a cornerstone of loop quantum gravity. Recently, there have been many new ideas in this field, and I will review some of them. In particular, after a brief description of the main structures and results of quantum geometry, I review a new description of the quantized geometry in terms of polyhedra, new results on the volume operator, and a way to incorporate a classical background metric into the quantum description. Finally I describe a new type of exponentiated flux operator, and its application to Chern-Simons theory and black holes. 
\end{abstract}

\section{Introduction}
\label{se_intr}
Loop quantum geometry is based on a canonical quantization of an SU(2) connection $A$ and its canonical conjugate $E$. $E$ encodes the intrinsic spatial geometry of a spatial slice 
$\Sigma$, whereas $A$ contains the extrinsic geometry of $\Sigma$ in space-time. The quantum theory of these fields -- before the implementation of the quantum version of Einstein's equations -- is called \emph{quantum geometry}. It is a cornerstone of loop quantum gravity (LQG). The basic properties of this quantum theory \cite{Rovelli:1989za,Rovelli:1994ge,Ashtekar:1995zh,Ashtekar:1996eg,Lewandowski:1996gk} have been studied a long time and are now well understood. Nevertheless, there has been a lot of activity in this area recently. While our basic understanding has not changed, new aspects have been discovered and there is a much refined understanding of many technical problems. Some of these developments have the potential to change the way we think about and approach the big challenges that confront LQG, such as a deeper understanding of the dynamics and the contact with observations. 

I have picked four areas of development for a review, namely 
\begin{enumerate}
	\item U(N) structure on intertwiner spaces and interpretation in terms of polygons and twisted geometries \cite{gr-qc/0501075,Freidel:2009ck,Freidel:2010aq,Freidel:2010tt,Rovelli:2010km,Bianchi:2010gc}.
  \item Sign factors and volume spectra \cite{Brunnemann:2007ca,Brunnemann:2010yv}.
  \item Quantum geometry with a classical spatial background \cite{Koslowski:2007kh,Sahlmann:2010hn}.
	\item New flux operators and their application to Chern-Simons theory and the description of black holes \cite{Sahlmann:2010bd,Sahlmann:2011uh,Sahlmann:2011xu,Sahlmann:2011rv}.
\end{enumerate}
These are not meant to be the most important ones. Rather, the idea was to give both examples of technical and of conceptual progress, which happened roughly speaking within the last two years. In the end, there is also a personal flavor to this selection, as I have worked on two of the topics myself. Thus there are important developments at least partially related to quantum geometry which I will not mention further in this report. For example, there are new results on the connection of the formalism of quantum geometry in symmetry reduced models to the full theory \cite{Fleischhack:2010zt,Brunnemann:2010qk}. Moreover, I will only mention the description of classical and quantum geometry in terms of SU(2) spinors \cite{Freidel:2010tt,Borja:2010rc,Livine:2011gp,Livine:2011vk,Livine:2011zz} in passing.
A very important topic that will be left out are the new insights regarding how the 3d and the 4d picture can come together \cite{Engle:2007wy,Ding:2009jq,Ding:2010ye},\cite{Alexandrov:2010pg}, \cite{Geiller:2011cv,Geiller:2011bh}. Similarly I will not consider the exciting developments on quantum geometry in higher dimensions \cite{Bodendorfer:2011nv,Bodendorfer:2011nw,Bodendorfer:2011nx,Bodendorfer:2011pa}.

\section{Quantum geometry -- a reminder}
\label{se_quan}
The basic variables that encode the quantized geometry are Wilson-loop and flux-like variables
\begin{equation}
W^{(j)}_\alpha=\text{tr}\left(\pi_j\left(P\exp\oint_\alpha A\right)\right),\qquad E_{S,f}=\int_S f^I \Sigma_{I}, \qquad \Sigma_I=\epsilon_{abc}E^a_I\, dx^b\wedge dx^c.
\label{eq:wilson}
\end{equation}
$\alpha$ is an oriented loop $j$ an SU(2) irrep, $S$ an oriented surface, and $f$ a smearing function taking values in su(2)$^*$. These variables are chosen because of their simple transformation properties under spatial diffeomorphisms and gauge transformations. In the quantum theory they are promoted to operators. Their commutation relations turn the set of operators into an algebra $\hfalg$, the \emph{holonomy-flux algebra}. The product of the operators $W^{(j)}_\alpha$ is determined by the pointwise product of the corresponding functionals, and hence in part by SU(2) representation theory. For example 
\begin{equation}
(W^{(j)}_\alpha)^2=\sum_{k=0}^{2j} W^{(k)}_\alpha. 
\end{equation}
A second class of relations in $\hfalg$ encodes the canonical commutation relations between $A$ and $E$. For example, for \includegraphics[width=5mm]{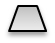} and \includegraphics[width=3mm]{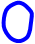} having a single transverse intersection of positive orientation \includegraphics[width=6mm]{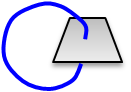} at a point $p$, 
\begin{equation}
[E_{\includegraphics[width=5mm]{box.png},f},W^{(j)}_{\includegraphics[width=3mm]{loop.png}}]=8\pi \beta \lp f^I(p) W^{(j)}_{{\includegraphics[width=3mm]{loop.png}} I},
\end{equation}
where ${W^{(j)}}_{{\includegraphics[width=3mm]{loop.png}} I}$ is an element of $\hfalg$ corresponding to an insertion to an element of su(2) in the functional \eqref{eq:wilson}. A third class of relations makes $\hfalg$ into a Lie algebra. For example 
\begin{equation}
[[E_{S,f},E_{S',f'}],W^{(j)}_\alpha]=(8\pi \beta \lp)^2 f^I(p)f^J(p)W^{(j)}_{\alpha IJ},
\label{eq:nc}
\end{equation}
where we have assumed that $S$ and $S'$ intersect each other and $\alpha$ in a point $p$. Again $W^{(j)}_{\alpha IJ}$ is an element of $\hfalg$ corresponding to an insertion of elements of su(2).

\begin{figure}
\centerline{\includegraphics[width=11cm]{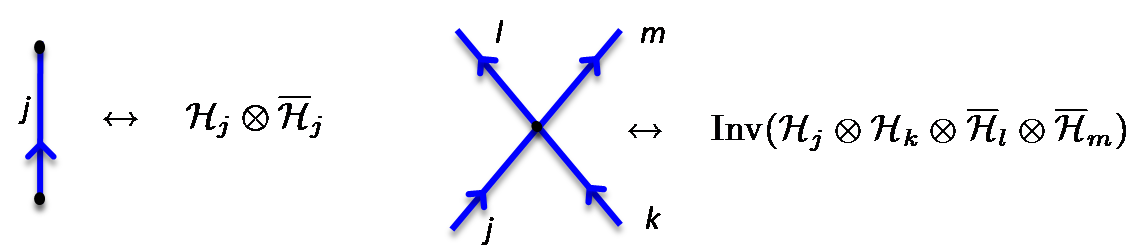}}%
\caption{Decomposition in to invariant spaces wrt.\ the operators $\vec{J}_e$. $\mathcal{H}_j$ stands for the spin $j$ representation of SU(2).}%
\label{fi_hspace}%
\end{figure}

An important representation for $\hfalg$, the \emph{Ashtekar-Lewandowski representation}. It is the unique representation invariant under spatial diffeomorphisms.  In it, the $W$ operators act as creation operators, generating states as excitations over a ground state $\ket{}$. An orthonormal basis in the corresponding Hilbert space $\mathcal{H}_\text{AL}$ can be labeled by \emph{spin networks}, diagrams that consist of a graph in $\Sigma$ and a labeling of graph edges with su(2) irreps and of graph vertices with invariants in the tensor product of the representations of edges attached to the respective vertex. There is a decomposition into orthogonal subspaces $\mathcal{H}_\gamma$ labeled by graphs $\gamma$. A finer decomposition into sub-Hilbert spaces can be made using the spin network diagrams. The rules are given in fig.\ \ref{fi_hspace}. 
The full Hilbert space also has an analytic description, $\mathcal{H}_\text{AL}=L^2(\overline{\mathcal{A}}/\overline{\mathcal{G}}, \text{d}\mu_{\text{AL}})$, as a space of square-integrable distributional connections modulo gauge transformations. 

In $\mathcal{H}_\text{AL}$, one can find operators corresponding to geometric quantities such as area, volume, length, or angle, in terms of the operators $J_e$ representing SU(2) at the end of the edges (see fig.\ref{fi_hspace}). As an example, consider the area operator. In a situation where a single simple loop intersects a surface, \includegraphics[width=6mm]{loopbox.png}, one has 
\begin{equation}
\widehat{A}_{\includegraphics[width=5mm]{box.png}}\ket{{\includegraphics[width=3mm]{loop.png}}_j}= 8\pi\beta l_p^2\, \sqrt{j(j+1)}\ket{{\includegraphics[width=3mm]{loop.png}}_j}.
\end{equation}
Volume operators act in the intertwiner spaces at the vertices. Their action is much more complicated and their spectrum is not known. At the same time, volume operators are a crucial ingredient in the quantum dynamics of LQG. In the next two sections we will present new results that also shed light on the action of the volume operators.   
\section{New developments}
\label{se_newd}
\subsection{Polyhedral interpretation}
\label{se_poly}
It is an old idea to associate some kind of classical geometry to a spin network: Since the volume operator acts at the vertices, volume can be thought to reside there. The edges on the other hand carry area. Conventionally, the geometry is thought of as distributional, the metric being degenerate except for points on the spin network graph. But recently, a new geometric picture emerged, which is supported by some old result on the geometry of polyhedra. The idea is to envision the geometry associated to the spin network as given by a gluing of polyhedra, one for each vertex. 

To describe the aspects of this new picture in more detail, we start by reminding the reader that given a graph $\gamma$ the associated Hilbert space of spin network states has a decomposition  
\begin{equation}
\mathcal{H}_\gamma 
=\bigoplus\limits_{\{j_e| e\in \gamma\}}\bigotimes\limits_{v\in\gamma} \text{Inv}(j_e, e\text{ at } v) .
\end{equation}
Thus it is natural to investigate the geometric interpretation of the intertwiner spaces associated to the vertices. The first to describe the geometries associated to the states in this Hilbert space in detail were Livine and Freidel \cite{Freidel:2009ck}, based on earlier work by Livine and Girelli \cite{gr-qc/0501075} which described an U($n$) structure on the $n$-valent intertwiner space
\begin{equation}
\mathcal{H}_n=\bigoplus\limits_{(j_1,j_2, \ldots j_n)} 
\text{Inv}(j_1,j_2, \ldots j_n).
\end{equation}
This action is obtained by expressing the generators of the SU(2) gauge rotations at a given vertex by two sets of annihilation and creation operators, the so-called \emph{Schwinger representation}. For one copy of SU(2), this reads 
\begin{equation}
J_z=\frac{1}{2}(a^\dagger a-b^\dagger b), \quad 
J_+=a^\dagger b, \quad
J_-=ab^\dagger.
\end{equation}
Now, with $n$ copies of SU(2) one has 2n sets of annihilation and creation operators, and 
the combination 
\begin{equation}
E_{ij}:= a^\dagger_ia_j+b^{\dagger}_ib_j
\end{equation}
commutes with global SU(2) rotations and has the commutation relations of U($n$) generators. Moreover, it preserves the subspace of intertwiners with a fixed sum 
$\sum_ij_i$ of incoming spins. this can be interpreted as the \emph{preservation of area}. Hence Livine and Freidel interpret the states in $\mathcal{H}_n$ as those 
of a quantized sphere, and the U($n$) action as discretized action of diffeomorphisms. The U($n$) structure is also useful for practical purposes, for example for determining the dimensions of various subspaces of the intertwiner space, and for constructing \emph{intertwiner coherent states} \cite{Freidel:2010tt}, a new type of coherent states contained in $\mathcal{H}_n$. The latter arise from coherent states for the harmonic oscillators in the Schwinger representation, and a suitable group averaging. Using the U(n) structure, one can also introduce spinorial coordinates in the space of intertwiners which give a wholly new perspective \cite{Freidel:2010tt,Borja:2010rc,Livine:2011gp,Livine:2011vk,Livine:2011zz}. 

Now one can also consider how to combine the chunks of quantized space. In \cite{Freidel:2010aq} Freidel and Speziale call the result \emph{twisted geometries}.  
\begin{figure}
\centerline{\includegraphics[width=7cm]{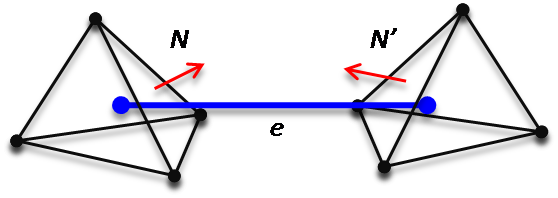}}%
\caption{The gluing of two flat tetrahedra dual to two vertices connected by an edge.}%
\label{fi_glueing}%
\end{figure}
$\mathcal{H}_e=L^2(SU(2))$ can be understood as carrier of a quantization of a phase space T$^*$SU(2) associated to the edge. But what is its interpretation in terms of physical quantities? The classic answer \cite{Thiemann:2000bw} is directly in terms of the fields $(A,E)$ used in LQG: Points in the phase space can be parametrized by $(h_e,\int_{S_e} E)$, where $h$ is the holonomy along $E$ and $S_e$ is a fixed surface transversal to $e$. The surfaces around a vertex form a curved polyhedron.
\cite{Freidel:2010aq, Rovelli:2010km} suggest to parametrize the phase space in a way that uses the geometry of two flat polyhedra glued together, possibly with extrinsic curvature. A point is given by $(N,N',j,\xi)$, where $N,N'$ are the unit normal vectors of the surfaces involved, and $j$ describes their area, see fig.\ \ref{fi_glueing}. The parameter $\xi$ is carrying the remaining information about the parallel transport from one polyhedron to the next. An interesting point, corresponding to an earlier result \cite{arXiv:0807.2806}, is that the geometries obtained in this way are twisted in the sense that the shapes of the glued triangles do not necessarily match.   

The picture is rounded off by the following observations \cite{Bianchi:2010gc}: according to a theorem by Minkowski, there is a 1-1 correspondence between sets $\{a_i\}_{i=1...n}$, $\{N_i\}_{i=1...n}$ of positive real numbers and unit vectors in $\R^3$ satisfying the closure relation 
\begin{equation}
\sum_i N_i a_i=0,
\label{eq:closure}
\end{equation}
and the equivalence classes of convex polyhedra under rotation. The $a_i$ correspond to the areas of the faces, the $N_i$ to their normal vectors under this correspondence. The space of this data carries a natural symplectic structure due to Kapovich and Millson, and its quantization can be precisely related to the structures at an $n$-valent vertex in LQG. To each edge $e$, one has a dual face whose area is given by the operator $|\vec{J}_e|$ and and whose product of area and normal vector corresponds to $\vec{J}_e$. The closure condition \eqref{eq:closure} is then just gauge invariance at the vertex, and the intertwiners acquire an interpretation as the quantum states of a convex polyhedron. This has very interesting consequences for the quantum black hole \cite{arXiv:1011.5628} and for the volume operator \cite{arXiv:1102.5439}. 

\subsection{Sign factors and volume spectra}
\label{se_volu}
The volume operators are a crucial ingredient in the quantum dynamics of LQG, but their action and spectrum is very complicated. Given a vertex $v$ one defines the operators  
\begin{equation}
\widehat{q}_{e,e',e''}=\text{sign}(\det(\dot{e},\dot{e}',\dot{e}''))\epsilon^{IJK} J_{I}^eJ_{J}^{e'}J_K^{e''},
\qquad 
\widehat{q}_v=\sum_{e,e',e'' \text{ at }v}\widehat{q}_{e,e',e''}
\end{equation}
The volume operator at the vertex is then given by 
\begin{equation}
\widehat{V}=\left(|\sum_v \widehat{q}_v|\right)^{1/2}
\quad\text{ or }\quad
\widehat{V}=\left(\sum_v |\widehat{q}_v|\right)^{1/2},
\end{equation}
depending on which version, Ashtekar-Lewandowski (first expression), or Rovelli-Smolin (second one) one considers. A particular complication of the first version version are the sign factors in the definition of $\widehat{q}_{e,e',e''}$, since these can have substantial influence on the spectrum for vertices of valence higher than three. While it may seem at first that arbitrary sign combinations may occur when letting $e,e',e''$ range over all the triples of edges at a given vertex, Brunnemann and Rideout \cite{Brunnemann:2007ca} have observed that by no means all sign combinations can actually be realized by configurations of tangent vectors in $\R^3$. How to effectively characterize the realizable sign configurations has been a question that has hampered the numerical analysis of the volume spectrum for some time. Recently however, Brunnemann and Rideout have found in matroid theory the tool to answer all the questions pertaining to these signs \cite{Brunnemann:2010yv}. Matroids are combinatorial structures generalizing linear dependencies in a real vector space, and \cite{Brunnemann:2010yv} describes in detail how they can be used to describe the diffeomorphism  invariance classes of tangential structure of vertices. 

\begin{figure}[htbp]
\center
\psfrag{frequency}{\#eigenvalues}
\psfrag{eigenvalue}{volume in $\ell_{\text{Planck}}^3$}
\includegraphics[width=7.5cm]{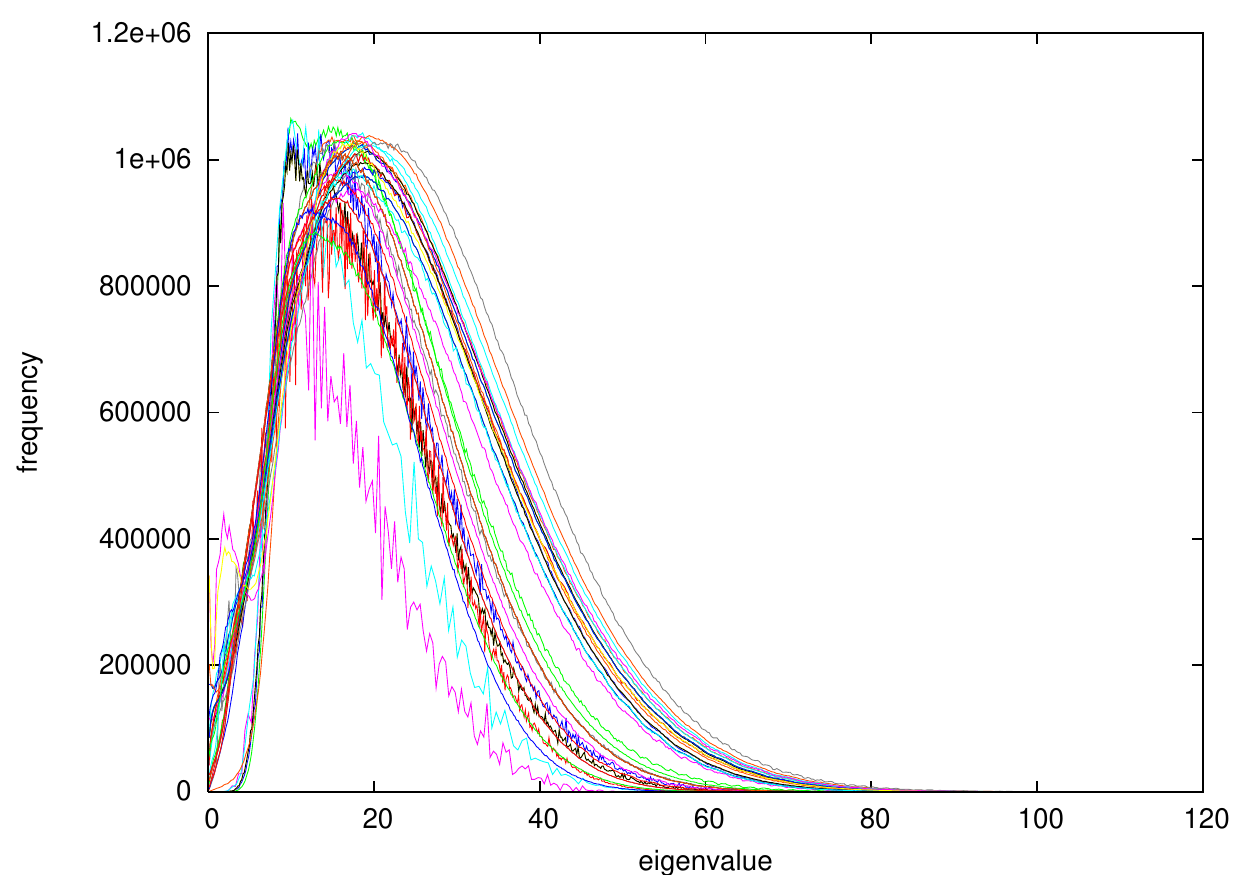}
\psfrag{frequency}{\#eigenvalues}
\psfrag{eigenvalue}{volume in $\ell_{\text{Planck}}^3$}
\includegraphics[width=7.5cm]{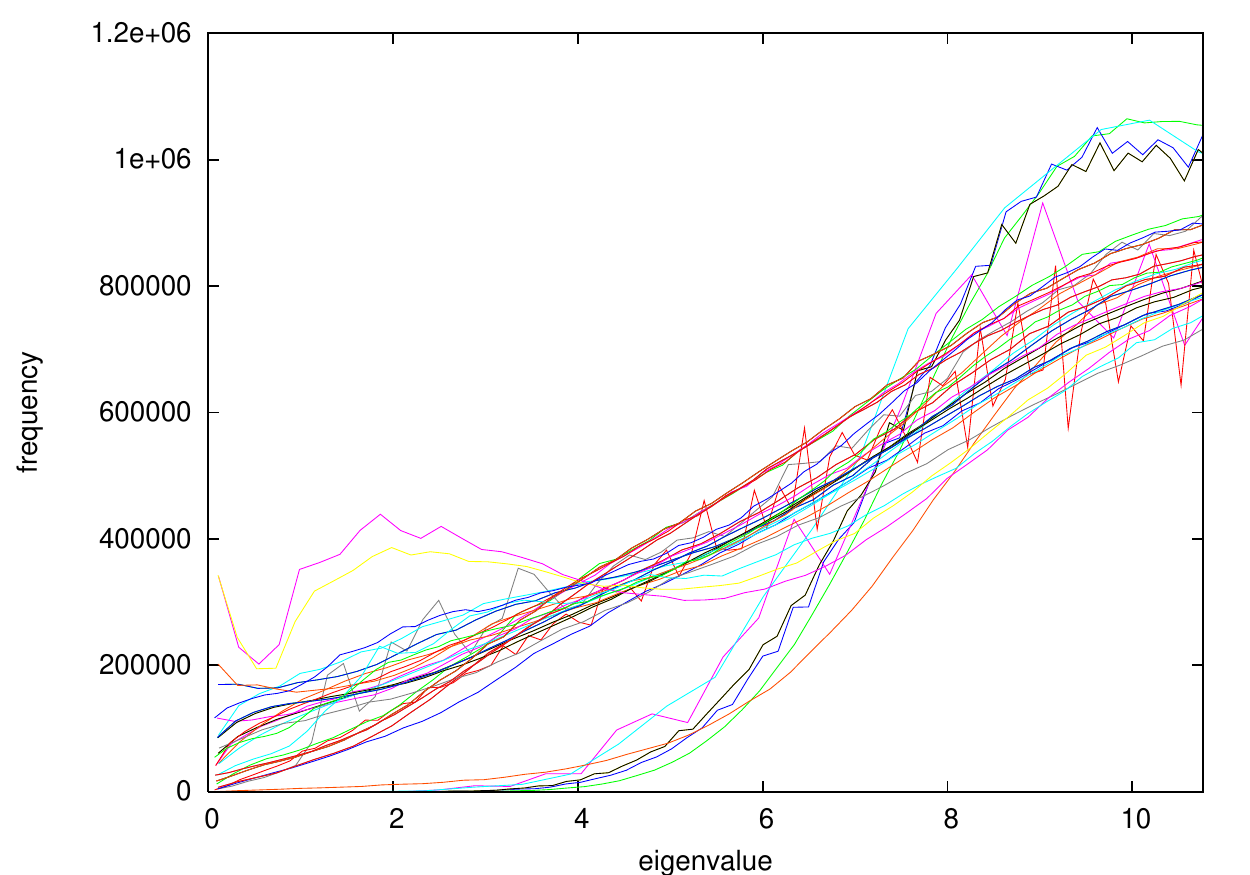}
\caption{From \cite{Brunnemann:2010yv} with kind permisson: Histograms for each sign configuration at the
 6-vertex, with incoming spins up to $j_\text{max}=13/2$.  The right figure shows the small eigenvalue region.}
\label{6v_sci_hist}
\end{figure}

Using these results together with some analytic ones \cite{gr-qc/0405060}, Brunnemann and Rideout have expanded the numerical study of the spectrum of the AL volume operator for various vertices, see for example  fig.\ \ref{6v_sci_hist}. 
It can clearly be seen that the sign combinations have a drastic effect on the eigenvalue density, with some combinations showing a \emph{volume gap}, whereas other combinations have high eigenvalue populations also for very small eigenvalues. These results will be quite important when the action of the Hamilton constraint (of which the AL volume operator is an important ingredient) is studied in detail. Additionally, as pointed out in \cite{Brunnemann:2010yv}, matroids can also be used to describe the combinatorics of edge composition in graphs, thus opening another potential application in LQG. 

\subsection{New representations}
\label{se_newr}
The AL representation of $\hfalg$ is uniquely singled out by its properties with respect to diffeomorphisms. But at least for some applications in situations where one considers highly excited states of the gravitational field, it may be interesting to also consider representations that are not invariant under diffeomorphisms. One class of such representations has been studied in \cite{Koslowski:2007kh, Sahlmann:2010hn}. Let $\bg{E}$ be a classical triad field. Then we can define a \emph{new}
representation $\pin$ on the AL Hilbert space $\hal$, by changing the action of the fluxes:
\begin{equation}
 \pin(E_{S,r})={X}_{S,r} +  \bg{E}_{S,r}\id, \quad \text{ with }  
\bg{E}_{S,r}=\int_S \bg{\Sigma}{}^I r_I.  
\label{eq:nflux}
\end{equation}
It is easily checked that this gives another representation of $\hfalg$. We
list some elementary properties: 
\begin{enumerate}
 \item The spectra of the fluxes have changed. If $\lambda$ is an eigenvalue
of $X_{S,f}$ then $\lambda + \bg{E}_{S,f}$ is an eigenvalue of $\pin(E_{s,f})$. 
In particular, the AL vacuum $\ket{}$ is now 
an eigenstate of $\pin(E_{s,f})$ with eigenvalue $\bg{E}_{S,f}$, which is 
non-vanishing in general.
\item The new representation is still cyclic, with $\ket{}$ as cyclic
vector.
\item The new representation is unitarily inequivalent to the AL representation.  

\item The standard kinematic representation can be viewed as a special case
of the new one, for $\bg{E}=0$. 
\end{enumerate}
To better interpret the new representations, it is very useful to consider the action of operators corresponding to geometric measures such as area and volume. It turns out that area and volume operators can be defined in the new representations with exactly the same regularization and quantization technique as in the standard representation, and the resulting operators have a very simple property:   
\begin{equation}
 V_{R}=V_R^{\vac}+ \bg{V}_R\id, \qquad A_S=A^{\vac}_S+\bg{A}_S\id,
\end{equation}
with $V^{\vac}, A^{\vac}$ the geometric operators in the vacuum representation,
and 
$\bg{V}_R,\bg{A}_S$ the classical values in the background geometry. 
This shows clearly that the new representations contain a background geometry given by the background field $\bg{E}$. 

Turn next to the gauge symmetries of LQG: The operators implementing diffeomorphisms and gauge transformations in the standard representation are still well defined and unitary in the new representation, but it can be easily checked that they do not implement the
algebra-automorphisms anymore. The problem is that operators on the Hilbert space of $\pin$ can not change 
change the background geometry. This can be remedied by going over to a large direct sum of Hilbert spaces. For a given background $\bg{E}$ one considers 
\begin{equation}
\label{eq_dirsum}
\mathcal{H}_{G(\bg{E})}:=\bigoplus_{\bg{\wt{E}}\in G(\bg{E})},
\mathcal{H}_{\bg{\wt{E}}}
\end{equation}
where $G(\bg{E})$ is the equivalence class of $\bg{E}$ with respect to
diffeomorphisms and gauge transformations. The Hilbert space is thus
effectively labeled by a spatial background metric modulo diffeomorphisms. 
The Hilbert space $\mathcal{H}_{G(\bg{E})}$ carries a representation of $\hfalg$, which is simply the direct sum of the representations on the
individual spaces. But now one can implement the diffeomorphisms and gauge
transformations unitarily \cite{Koslowski:2007kh,Sahlmann:2010hn}. 
The unitary implementation of the symmetries can now be used to find invariant states and Hilbert spaces by \emph{group averaging}, just as has been done for the vacuum representation \cite{Ashtekar:1996eg}. For example, one obtains diffeomorphism invariant functionals $(T_\alpha,\bg{E}|$ on cylindrical functions by averaging with respect to certain quotients of the diffeomorphism group.  
Similar results can be obtained for gauge transformations and, taking the semi-direct product of diffeomorphisms and gauge transformations, for bundle automorphisms. One thereby obtains the automorphism invariant Hilbert space $\mathcal{H}_\text{aut}$. Thus the quantum kinematics can be developed to exactly the same point for the new representations as for the AL representation. What is more, the latter appears simply as a special case of the constructions we sketched above. 

I should stress that, while $\mathcal{H}_{G(\bg{E})}$ is very large, this is partially remedied by the group averaging. Vectors 
$\ket{1,\bg{E}}$, $\ket{1,\bg{E'}}$ are mapped onto \emph{the same} vector in $\mathcal{H}_\text{aut}$. More generally   
\begin{equation}
(f,\phi \pone E|=(\phi^{-1}\pont f,E|.
\end{equation}
I should also say that it seems that there is no problem with operators on $\mathcal{H}_\text{aut}$. As an example consider the operator $V$ for the volume of the entire spatial slice. Also in the new representations, it commutes with all automorphisms. It thus defines an operator on $\mathcal{H}_\text{aut}$. Moreover, this operator acts in precisely the way one would expect. If $f$ is an eigenstate of $V^{\vac}$, with eigenvalue $\lambda$, then  $|f,\bg{E})$ is an eigenvector of $V$ with eigenvalue $\lambda+\bg{V}$. 
\subsection{Non-commuting fluxes, Chern-Simons theory and horizons}
\label{se_nonc}
I start by reminding the reader that the flux operators are mutually non-commuting in quantum geometry, see \eqref{eq:nc}. This means that there is some form of non-commutative geometry present in LQG. Recently, it turned out that this is not only a nuisance, but very useful in applications to Chern-Simons theory and black holes \cite{Sahlmann:2010bd,Sahlmann:2011uh,Sahlmann:2011xu,Sahlmann:2011rv}. The starting point are new gauge covariant exponentiated flux operators. Usually, the field $E$ is integrated over surfaces in the quantum theory, but it turns out that even the non-integrated quantum field makes sense as an operator valued distribution $\widehat{E}$. Then given a surface $S$ without self-intersections, and a system of paths that connect all points of the surface with some base-point on the edge of the surface, one defines   
\begin{equation}
W_S=\um_{2}
+ 8\pi i c \int_S  \aad_{h_s}(\widehat{\Sigma}(s))
+\left(8\pi i c\right)^2\int_{S^2} K_{s,s'}  
\aad_{h_s}(\widehat{\Sigma}(s))
\aad_{h_{s'}}(\widehat{\Sigma}(s'))
+\ldots
\end{equation}
Here, $K$ is an integration kernel that implements a \emph{surface ordering} and the holonomies $h_s$ connect the point $s$ on the surface with the base point via the path system. 
There are two problems with the above definition. The first is that consecutive actions of $\widehat{E}^a(p)$ give rise to delta distributions that are concentrated precisely at the boundary of integration enforced by surface ordering, and a prescription for the evaluation of these has to be adopted. This can be resolved via a standard regularization. The second problem is that the operators $\widehat{E}_I(p)$ at a fixed point $p$ do not commute, therefore there is an ordering ambiguity inherent in the above definition. This ambiguity can be fixed using the \emph{Duflo map}. This is a quantization map $\Upsilon$ from the free algebra of symbols $\{E_I\}_I$ with the Poisson bracket $\{E_I,E_J\}=f_{IJ}{}^KE_K,$
($f$ being the structure constants of a semisimple Lie-algebra $\mathfrak{g}$) into the universal enveloping algebra $U(\mathfrak{g})$, extending the map  $E_I\mapsto X_I$ on generators. The defining property of $\Upsilon$ is that it is an \emph{algebra isomorphism} between the invariant subspaces under the action of the corresponding Lie group $G$. 
$\Upsilon$ is an improved version of symmetric quantization $\chi$, 
\begin{equation}
\Upsilon = \chi\circ j^{\frac{1}{2}}(\partial), 
\label{eq:duflo}
\end{equation}
where $j^{\frac{1}{2}}(\partial)$ is a differential operator that can be obtained by inserting derivatives $\partial^I$ into the following function on $\mathfrak{g}$:
\begin{equation}
j^\frac{1}{2}(x)
=\det{}^\frac{1}{2}\left(\frac{\sinh\frac{1}{2}\aad x}{\frac{1}{2}\aad x}\right)
= 1 +\frac{1}{48} \norm{x}^2 +\ldots,
\label{eq:j12}
\end{equation}
with $\norm{x}^2=\tra(\aad_x^2)$ the square of the Cartan-Killing norm of $G$.
For $G$=SU(2) on finds for example $\Upsilon (\norm{E}^2)= \Delta_{\text{SU(2)}} +\frac{1}{8}\one,$
where $\Delta_{\text{SU(2)}}$ is the Laplacian. I should also point out that the Duflo map has recently been used in \cite{arXiv:1105.0439} and it shows up in the context of the \emph{flux representation} \cite{hep-th/0502106,hep-th/0601004,arXiv:1004.3450}. Its use has also been advocated, in a slightly different context, in \cite{hep-th/0405036}.  
Using $\Upsilon$ to order powers of $E$ the action of the operator $W_S$ becomes well defined.  
The operators $W_S$ satisfy some useful relations. For $S$, $S_1$, $S_2$ surfaces, $S_1+S_2$ the disjoint union of the latter, 
\begin{equation}
W_{S_1+S_2}=W_{S_1}W_{S_2}, \qquad W^\dagger_S=W_{-S}, \qquad U_g\, W_S\, U_g^{-1}= g(b) W_S g(b)^{-1},
\label{eq:prod}
\end{equation}
where $-S$ is $S$ with opposite orientation, $U_g$ is the unitary corresponding to a gauge transformation $g$, $b$ is the base-point of $S$, and paths on $S_1$, $S_2$, $S_1+S_2$ have been chosen in a compatible way. 
Moreover, for SU(2) and spin nets $\ket{\njp}$ intersecting $S$ transversally once with spin $j$, one finds 
\begin{equation}
\tra_j\left(W_{S}\right)\ket{\njp}
=\frac{\sin\left[\pi c (2j+1)(2j'+1)\right]}{\sin\left[\pi c(2j'+1)\right]}\ket{\njp}.
\label{eq:ev}
\end{equation}
Finally, states $\ket{\nj}$ are not eigenstates of the operator valued matrices $W_S$. Rather,      
\begin{equation*}
W_S\ket{\nj}= c_1(j) \um_{2\times 2} \ket{j}+ c_2(j) \ket{\njd}, \qquad \ket{\njd}= \aad_{h_p}(T_I)\kappa^{IJ} \widehat{E}_J(p)\ket{\nj},
\end{equation*}
where $c_1(j),c_2(j)$ are constants that are complicated to compute explicitly, and $\ket{\njd}$ carries a new vertex at $p$ and a link between $p$ and the base point of the loop along the path system on $S$. 
\subsubsection{Application to Chern Simons theory}
Since the work of Witten \cite{Witten:1988hf}, it is well known that path integral expectation values of holonomy traces in CS theory are related to link invariants. For $G=$SU(2) and traces in the defining representation one obtains the \emph{Kauffman bracket}, with the conventional variable $A$ replaced by $q^{1/4}$, $q=\exp(2\pi i/k)$.\footnote{In fact, with CFT methods on finds $q=\exp(2\pi i/(k+2))$ \cite{Witten:1988hf}, but we do not see the shift of the level with our method.} One can reproduce some of these results using the operators $W_S$. The key is that 
\begin{equation}
-\frac{8\pi  i}{k}\epsilon_{abc}\frac{\delta}{\delta A_c}e^{i S_{\text{CS}}[A]}= F_{ab}e^{i S_{\text{CS}}[A]}
\label{eq:func}
\end{equation}
for SU(2) CS theory with level $k$, and same, up to a numerical factor, for other gauge groups. The non-Abelian version of Stokes' theorem suggests that holonomy functionals can be replaced by the new flux operators under the path integral. To make this reasoning explicit, let $S$ be a smooth, oriented, simply connected surface, $\rho$ some representation of the structure group $G$, and and $L$ be some functional of $G$-connections. Then formally  
\begin{equation}\begin{split}
\expec{L \tra_\rho(h_{\partial_S})}&=\int_{\conn} L[A] \tra_\rho(h_{\partial_S})[A] e^{i
S_{\text{CS}}[A]}\, \dm [A]=\int_{\conn}L[A] \tra_\rho(W_S)e^{i S_{\text{CS}}[A]}\, \dm [A]\\
&=\int_{\conn} (\tra_\rho(W_{-S})L)[A]
e^{i S_{\text{CS}}[A]}\, \dm [A]
=\expec{(\tra_\rho (W_{-S})L)}.
\end{split}
\label{eq:logic}
\end{equation}
While the manipulations under the path integral are formal, taking the first and the last line gives an equality in which all objects are defined, at least as long as $S$ has no self-intersections. Consider $G$=SU(2) and chose $c=1/k$ with $k$ integer \cite{180552,MIT-CTP-964} in the definition of $W_S$. 
Let $H^+(j_1,j_2)=\tr_{j_1} (h_{\alpha_1})\tr_{j_2} (h_{\alpha_2})$ 
denote the right handed Hopf link spin network. 
Using the above strategy and the eigenvalues \eqref{eq:ev}, one finds 
\begin{equation}
\expec{H^+(j_1,j_2)}_\text{CS}=\tr_{j_1}(g_{j_2})\tr_{j_2}(g_0)=\frac{\sin\left[\frac{\pi}{k}(2j_1+1)(2j_2+1)\right]}{\sin\left[\frac{\pi}{k}\right]},
\label{eq:verlinde}
\end{equation}
thus reproducing the known values for the Kauffman bracket and its generalization \cite{Martin:1989ub}. 
Note that these numbers are important in related contexts: They are equivalently given by the trace of the square of the R-matrix of U$_q$(su(2)) on $j_1\otimes j_2$ or, up to normalization, by the Verlinde coefficients \cite{Verlinde:1988sn} in the SU(2)$_k$ WZW model of conformal field theory. 
Similar results can be obtained for other gauge groups \cite{Sahlmann:2011uh}.
\subsubsection{Application to Black holes}
The quantization of an isolated horizon (IH) is a remarkable success of LQG (see \cite{arXiv:1112.0291} for a recent overview). However, it is only an effective description, in the sense that it uses a number of elements that are not intrinsic to the formalism of LQG. This can be partially overcome using the operators $W_S$. 
For spherically symmetric IH, the pullback to the IH, of the dynamical fields satisfy the horizon condition \cite{Engle:2010kt}
\begin{equation}
\pb{F}(A)=-\frac{\pi(1-\beta^2)}{a_H}\pb{\Sigma}(E).
\label{eq:horizon}
\end{equation}
In the quantum theory, we thus call a surface $H$ a type I quantum horizon, if the surface ordered exponential of the above equation, $\tr{h}_{\partial S}\Psi= \tr{W_S}\Psi$,  
is valid for all surfaces $S$ lying entirely within $H$. Here, the operators $W_S$ are evaluated with $c=-\pi\beta(1-\beta^2)\lp^2/2a_H$. Solutions $\Psi$ of this equation do not exist in the standard representation of LQG, but arguably in other representations. A spin network $\ket{\psi}$ determines puncture data $\mathcal{P}=\{(p_1,j_1,m_1),(p_2,j_2,m_2),\ldots (p_N,j_N,m_N)\}$
given by points $p_i$ on $H$ irreducible representations $j_i$ and and magnetic quantum numbers $m_i$. $\mathcal{P}$ defines a functional on functions of traces of loops that encircle at most one puncture:
\begin{equation}
\mu(\prod_i \tr_{k_i}(h_{\alpha_i})):= \prod_i \tr_{k_i}(g_{j_i}).  
\label{eq:su2def}
\end{equation}
This functional is positive and consistent with all the relations among these traces. It is however not yet clear whether it extends all holonomy functionals on $H$. 
The resulting horizon theory would seem to reproduce many of the results that have been obtained earlier \cite{Engle:2009vc,Engle:2011vf}, up to the fact that the fluxes $E$, pulled back to the horizon are still a-priori independent degrees of freedom describing the intrinsic geometry of $H$.  
\begin{samepage}
\section{Concluding remarks}
\label{se_conc}
I hope that it has become clear that quantum geometry still intrigues with new facets and discoveries. All of this is potentially relevant for physics: The polyhedral picture is already applied to give new forms of dynamics and for a better understanding of the quantum theory of black hole horizons. The detailed knowledge of spectrum and properties of the volume operator are relevant for the implementation of the Hamilton constraint, and the representations with classical background may be used to construct effective versions of the dynamics. The new flux operators have already proven useful in the description of black hole horizons. But since these are new developments, other applications are bound to emerge. I hope that this short overview can be helpful in this respect. 
\ack 
I thank the organizers of \emph{Loops 11} for making it such a nice and productive event, and for the opportunity to speak. I gratefully acknowledge partial support through the Spanish MICINN Project No. FIS2008-06078-C03-03.
\end{samepage}

\end{document}